# Polariton spectrum in nonlinear dielectric medium

**Igor V. Dzedolik, O. Karakchieva**

*Taurida National V. I. Vernadsky University, 4, Vernadsky Avenue,*

*95007, Simferopol, Ukraine,*

*dzedolik@crimea.edu , karakchieva@crimea.edu*

We obtain theoretically the phonon-polariton spectrum in nonlinear dielectric medium with the third order Kerr-type nonlinearity. We investigate the dependence of number of the polariton spectrum branches on the intensity of electromagnetic field and demonstrate that the appearance of new branches located in the polariton spectrum gap is caused by the influence of dispersion of the third order dielectric susceptibility at intensity field in the medium. The modulation instability of new branch waves leading to appearance of cnoidal waves and solitons.

### 1. INTRODUCTION

The velocity of electromagnetic wave propagating through a medium decreases in comparison with velocity in vacuum due to the interaction of the wave and ions of medium. The interaction produces the moving of electron shells and ion nuclei. At the linear approximation it is sufficient to take into account the dipole ion excitation (electron shell and nucleus) by electromagnetic field. The spectra of polaritons that are the quanta of mixed states of electromagnetic wave and dipole excitations of the dielectric media were obtained firstly by Tolpygo using quantum approach [1] and by Huang using classical approach [2]. This phonon-polariton spectrum has two branches and band gap between them, where the crystal is not transparent.

The polaritons are efficiently generated when the frequency of electromagnetic wave is close to the crystal lattice resonance frequency lying in terahertz range $\sim 10^{12} s^{-1}$ or to the electron resonant frequency lying near the infrared or optical ranges $\sim 10^{14} s^{-1}$ [1 - 4]. It leads to enhancement of vibration amplitude, i.e. to the nonlinear oscillations of ions and electrons. Nonlinear effects also appear in generation of polaritons out of these ranges, if the nonlinear electron response of medium arises due to intensive electromagnetic wave scattering [5 - 8]. In these cases some new branches of phonon-polariton spectrum appear [5, 8].

Polariton spectra were obtained in multilayered nonlinear medium in Ref. [8], where the authors showed that the number of spectral branches was determined by the wavevector value, phonon frequency, thickness and nonlinearity of each layer. A new technique of measuring of the reflection in terahertz range by phonon-polariton wave was proposed in the Ref. [9]. The phonon-polariton propagation in one-dimensional piezoelectric superlattice was studied in Ref. [10]. The properties of phonon-polariton waves, the masses of polaritons and the dependence of polariton spectrum on intensity of the electromagnetic field and intensity of the external electric field were investigated in nonlinear dielectric mediums and waveguides in Refs. [11 - 13]. The experimental observations of interference between atomic spin coherence and optical field generating polariton waves in controllable time-delayed beamsplitter with dynamically tunable splitting ratio were described in Ref. [14]. Nonlinear waves described by the fifth order dispersive equation have been numerically investigated in Ref. [15]. The band-gap solitons in Kerr-type nonlinear media with periodic in space modulation of potential were also researched and the instability of the eigenfrequencies around the middle of the band-gap was found in Ref. [16]. The above mentioned papers are devoted to the analysis of polariton properties in nonlinear media and illustrate common interest of researches in this field of physics. The phonon-polariton properties in dielectric medium are important in the design of controllable filters, all-optical logic gates, signal delay lines, time-delayed beamsplitters and other devices of microwave and optical circuits [6 - 10, 14].

In our paper we theoretically investigate the properties of phonon-polariton spectrum in nonlinear dielectric medium with the third order susceptibility, i.e. the Kerr-type nonlinear medium. The dispersion relation of frequency $\omega(k)$ on the wavevector $k$ in nonlinear medium, $D(\omega, k, E_a) = 0$, is modified by the field amplitude $E_a$. Thus we obtain the nonlinear dispersion relation similar to that in Refs. [17, 18].

The nonlinear polaritons represent the bound states of photons and phonons in the dielectric medium that depend on the intensity of electromagnetic field. We show how intensity of electromagnetic field influences the number of polariton spectrum branches. The enhancement of electromagnetic field intensity leads to increasing of nonlinear response of dielectric medium and it results in appearance of the additional branches in the polariton spectrum gap. To demonstrate this effect we have obtained the phonon-polariton spectra in nonlinear medium considering the dispersion of the third order nonlinear susceptibility for the first harmonic of wave frequency.

## 2. THEORETICAL MODEL

Consider the theoretical model of the classical electromagnetic field interaction with the ions forming a crystal lattice. If the electromagnetic wave propagates in the dielectric crystal we can describe this process using the following:

1) the equation of ion motion in a unit cell of crystal lattice

$$m_{eff}\frac{d^2\mathbf{R}}{dt^2} + m_{eff}\Gamma\frac{d\mathbf{R}}{dt} + \nabla_R U_R = e_{eff}\left(\mathbf{E} + \frac{1}{c}\frac{d\mathbf{R}}{dt}\times\mathbf{B}\right), \qquad (1)$$

where $e_{eff}, m_{eff}$ are the effective ion charge and mass in the lattice cell, $\mathbf{R} = \mathbf{r}_+ - \mathbf{r}_-$ is the displacement vector of positive and negative lattices of ions, $U_R = (q_{1R}/2)R^2 + (q_{2R}/3)R^3 - (q_{3R}/4)R^4$ is the potential energy of ions, $\Gamma$ is the damping factor;

2) the equation of outer shell electron motion in the ion

$$m\frac{d^2\mathbf{r}}{dt^2} + m\Gamma\frac{d\mathbf{r}}{dt} + \nabla_r U_r = -e\left(\mathbf{E} + \frac{1}{c}\frac{d\mathbf{r}}{dt}\times\mathbf{B}\right), \qquad (2)$$

where $U_r = (q_{1r}/2)r^2 + (q_{2r}/3)r^3 + (q_{3r}/4)r^4$ is the potential energy of electron;

3) the electromagnetic field equations

$$\nabla\times\mathbf{B} = c^{-1}(\dot{\mathbf{E}} + 4\pi\dot{\mathbf{P}}), \quad \nabla\times\mathbf{E} = -c^{-1}\dot{\mathbf{B}}, \qquad (3)$$

where $\mathbf{P} = e_{eff}N_C\mathbf{R} - eN_e\mathbf{r}$ is the polarization vector of medium, $N_C$ is the number of cells in the unit of volume, $N_e$ is the number of electrons in the unit of volume, $q_{jr}, q_{jR}$ are the phenomenological elastic parameters of the medium; the overdot denotes partial time derivative. In the system of equations (1) - (3) we take into account the bound of charges by electromagnetic field.

We can neglect the response of the magnetic component of the high-frequency electromagnetic field $|\mathbf{E}| \gg |c^{-1}(d\mathbf{R}/dt)\times\mathbf{B}|$, $|\mathbf{E}| \gg |c^{-1}(d\mathbf{r}/dt)\times\mathbf{B}|$ in the medium. Then we represent solutions of the motion equations (1) and (2) as the series where index of term is the order of infinitesimal $\mathbf{r} = \mathbf{r}_0 + \mathbf{r}_1 + \mathbf{r}_2 + \mathbf{r}_3$, $\mathbf{R} = \mathbf{R}_0 + \mathbf{R}_1 + \mathbf{R}_2 + \mathbf{R}_3$. If the electromagnetic field is harmonic $E \sim exp(-i\omega t)$, it is easy to obtain the polarization vector of medium by the method of sequential approximations [6].

Generally by this method we can obtain the polarization vector including the first, second and third harmonics

$$\mathbf{P} = \chi_1 \mathbf{E}_a \exp(-i\omega t) + \chi_{20} E_a \mathbf{E}_a + \chi_{22} E_a \mathbf{E}_a \exp(-i2\omega t) \\ + \chi_{31} E_a^2 \mathbf{E}_a \exp(-i\omega t) + \chi_{33} E_a^2 \mathbf{E}_a \exp(-i3\omega t),$$  (4)

where $\chi_1 = \dfrac{1}{4\pi}\left(\dfrac{\omega_e^2}{\tilde{\omega}_1^2} + \dfrac{\omega_I^2}{\tilde{\Omega}_1^2}\right)$, $\chi_{20} = \dfrac{1}{4\pi}\left(\dfrac{e\alpha_{2r}\omega_e^2}{m\omega_0^2\left((\omega_0^2-\omega^2)^2 + \omega^2\Gamma^2\right)} - \dfrac{e_{eff}\alpha_{2R}\omega_I^2}{m_{eff}\Omega_\perp^2\left((\Omega_\perp^2-\omega^2)^2 + \omega^2\Gamma^2\right)}\right)$,

$\chi_{22} = \dfrac{1}{4\pi}\left(\dfrac{e\alpha_{2r}\omega_e^2}{m(\tilde{\omega}_1^2)^2 \tilde{\omega}_2^2} - \dfrac{e_{eff}\alpha_{2R}\omega_I^2}{m_{eff}(\tilde{\Omega}_1^2)^2 \tilde{\Omega}_2^2}\right)$, $\chi_{31} = -\dfrac{1}{4\pi}\left(\dfrac{e^2\alpha_{3r}\omega_e^2}{m^2(\tilde{\omega}_1^2)^3(\tilde{\omega}_1^2)^*} + \dfrac{e_{eff}^2\alpha_{3R}\omega_I^2}{m_{eff}^2(\tilde{\Omega}_1^2)^3(\tilde{\Omega}_1^2)^*}\right)$

$\chi_{33} = -\dfrac{1}{4\pi}\left(\dfrac{e^2\alpha_{3r}\omega_e^2}{m^2(\tilde{\omega}_1^2)^3 \tilde{\omega}_3^2} + \dfrac{e_{eff}^2\alpha_{3R}\omega_I^2}{m_{eff}^2(\tilde{\Omega}_1^2)^3 \tilde{\Omega}_3^2}\right)$ are the linear and nonlinear susceptibilities of medium,

$\tilde{\omega}_1^2 = \omega_0^2 - \omega^2 - i\Gamma\omega$, $\tilde{\omega}_2^2 = \omega_0^2 - (2\omega)^2 - i2\Gamma\omega$, $\tilde{\omega}_3^2 = \omega_0^2 - (3\omega)^2 - i3\Gamma\omega$, $\tilde{\Omega}_1^2 = \Omega_\perp^2 - \omega^2 - i\Gamma\omega$,

$\tilde{\Omega}_2^2 = \Omega_\perp^2 - (2\omega)^2 - i2\Gamma\omega$, $\tilde{\Omega}_3^2 = \Omega_\perp^2 - (3\omega)^2 - i3\Gamma\omega$; $\omega_e^2 = 4\pi e^2 N_e m^{-1}$, $\omega_I^2 = 4\pi e_{eff}^2 N_C m_{eff}^{-1}$ are the electron and ion plasma frequencies; $\omega_0^2 = q_{1r} m^{-1}$ is the electron resonance frequency, $\Omega_\perp^2 = q_{1R} m_{eff}^{-1}$ is the resonance frequency of lattice; $\alpha_{2r} = q_{2r} m^{-1}$, $\alpha_{3r} = q_{3r} m^{-1}$, $\alpha_{2R} = q_{2R} m_{eff}^{-1}$, $\alpha_{3R} = q_{3R} m_{eff}^{-1}$; $\Gamma, q_{jr}, q_{jR}$ are the phenomenological parameters depending on linear and nonlinear properties of medium, including also electron $\omega_e$ and ion $\omega_I$ plasma frequencies. The susceptibility of the third order is positive $\chi_{31} > 0$, because $\alpha_{3R} = -q_{3R} m_{eff}^{-1} < 0$, $\alpha_{3r} = -q_{3r} m^{-1} < 0$ are negative.

### 3. NONLINEAR MEDIUM WITH LOCAL INVERSION CENTERS

We consider the application of our theory only in the medium with local centers of inversion, where the second-order susceptibility vanishes due to symmetry, i.e. in the medium with the third order susceptibility $\chi_{31}$. We represent the electromagnetic field as the set of plane waves and consider the interaction degenerates involving only at the first harmonic $\mathbf{E} = \mathbf{E}_a \exp(-i\omega t + i\mathbf{k}\mathbf{r})$. In this case the polarization vector (4) of the medium has the form $\mathbf{P} = \left(\chi_1 + \chi_{31} E_a^2\right)\mathbf{E}$.

We can eliminate the magnetic inductance vector **B** from the field equations (3) $\nabla\times\nabla\times\mathbf{E}+c^{-2}\varepsilon\ddot{\mathbf{E}}=-4\pi c^{-2}\ddot{\mathbf{P}}$, and get the set of algebraic equations

$$\left(k^{2}-c^{-2}\varepsilon\omega^{2}\right)\mathbf{E}_{a}=\mathbf{k}\left(\mathbf{kE}_{a}\right). \tag{5}$$

The permittivity of medium is defined by the expression $\varepsilon=1+4\pi\chi_{1}+4\pi\chi_{31}E_{a}^{2}$. In the expression for permittivity $\varepsilon$ both the electron and ion responses of the medium at electromagnetic field are considered.

We resolve the electric field vector at the transverse and longitudinal components $\mathbf{E}=\mathbf{E}_{\perp}+\mathbf{E}_{/\!/}$ relating to the wavevector **k**. Having assumed that the interaction of electromagnetic waves and charges in the medium occurs as affected transverse field component $\mathbf{E}_{\perp}$, that is $\mathbf{kE}_{\perp}=0$, we obtain from equation (5) the dispersion equation for nonlinear polaritons in the Kerr-type medium

$$k^{2}-c^{-2}\omega^{2}\left[1+4\pi\chi_{1}(\omega)+4\pi\chi_{31}(\omega)E_{a}^{2}\right]=0. \tag{6}$$

The equation (6) allows obtaining the nonlinear phonon-polariton spectrum $\omega=\omega(k,E_{a}^{2})$ in this medium, where the frequency $\omega$ (or energy $\hbar\omega$) of polariton depends on the wave density $\sim E_{a}^{2}$ of polariton wave.

## 4. POLARITON SPECTRUM IN THE NONLINEAR MEDIUM

The polariton spectrum depends on the density of electromagnetic field $\sim E_{a}^{2}$ in the Kerr-type nonlinear medium. In the linear medium at $4\pi\chi_{31}E_{a}^{2}\to 0$ the spectrum of polaritons has only three branches (two low frequency and one high frequency branches, Fig. 1a), and this result agrees with the deduction in [5]. In the nonlinear medium, for example at $4\pi\chi_{31}E_{a}^{2}=10^{-5}$, the polariton spectrum has nine branches (Fig. 1b). In this case the spectrum still has branches 1, 2, 3, but six new branches with numbers 4, 5, 6 and 7, 8, 9 appear. New branches 4 and 5 coincide with themselves and smoothly go up, but the branch 6 has a weak declination down. The branches 7, 8, 9 have the same behavior: the branch 7 has the weak declination down, and branches 8, 9 smoothly go up and completely coincide.

Polariton spectrum in the linear medium has only two gaps, but as the electromagnetic field density increases, the third gap appears: the first gap (between branch 2 and branches 4, 5, 6), the second gap (between branch 2 and branches 7, 8, 9) and the third gap (between a branch 3 and branches 7, 8, 9).

The appearance of new branches in the polariton spectrum is caused by the dispersion of the third order dielectric susceptibility of medium $\chi_{31}(\omega)$ at increasing of the electromagnetic field density $\sim E_a^2$ (see the expression (4)). In the linear medium the dispersion equation (6) has the sixth degree of the frequency $\omega$. In the nonlinear medium the dispersion equation (6) has the eighteenth degree of the frequency because the susceptibility of the third order possesses the dispersion. In other words the more high frequency polaritons of branches 3 (and 2) decay to three low frequency polaritons of branches 7, 8, 9 (and 4, 5, 6), therefore the new spectrum branches appear.

The spectrum curves have been obtained numerically by solving the dispersion equation (6). As the example of computing the nonlinear polariton spectrum we use the medium and wave parameters as $\Omega_\perp \sim 10^{13} s^{-1}$, $k = 3\times10^2...3\times10^3 cm^{-1}$ in the Fig. 1.

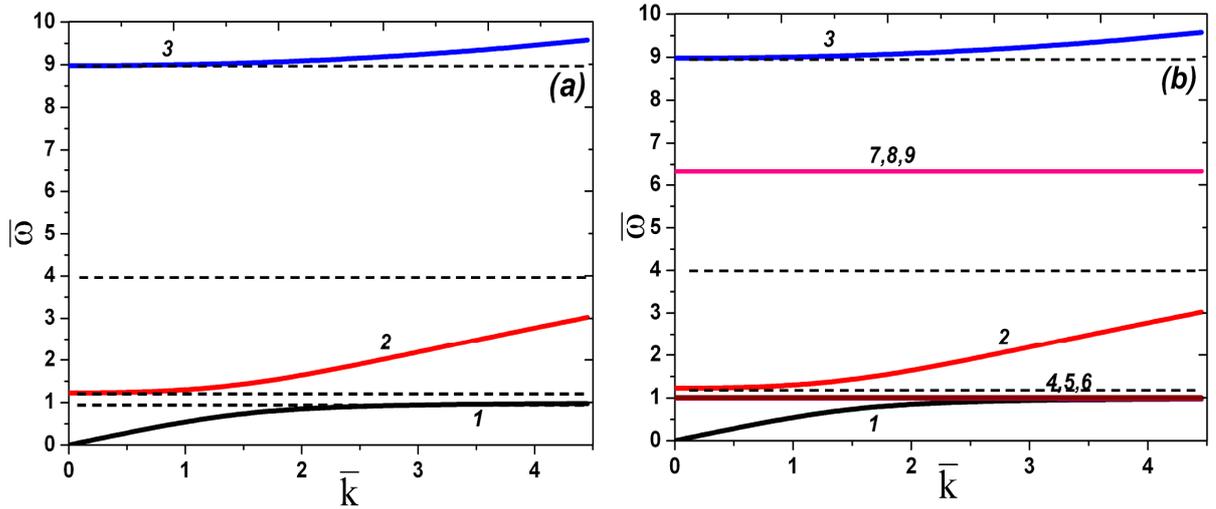

Fig. 1. The polariton spectra in nonlinear medium with dispersion of the third order susceptibility $\chi_3$: (a) at $4\pi\chi_{31}E_a^2 \to 0$; (b) $4\pi\chi_{31}E_a^2 = 10^{-5}$. Here $\Gamma = 0$, $\overline{\omega} = \omega/\Omega_\perp$ and $\overline{k} = ck/\Omega_\perp$ are dimensionless. The solid lines are the branches of polariton spectrum; the dashed horizontal lines are the edges of spectrum gaps.

## 5. WAVE INSTABILITY

We can investigate the wave instability of new spectrum branches by the equation obtained from the electromagnetic field equations (3) for transverse electric field in the considering case

$$\frac{\partial^2 E}{\partial z^2} - \frac{\varepsilon_1}{c^2}\frac{\partial^2 E}{\partial t^2} - \frac{4\pi\chi_{31}}{c^2}\frac{\partial^2 |E|^2 E}{\partial t^2} = 0, \tag{7}$$

where $\varepsilon_1 = 1 + 4\pi\chi_1 = \varepsilon_1' + i\varepsilon_1''$, $\chi_{31} = \chi_{31}' + i\chi_{31}''$, $\Gamma \neq 0$. Considering the perturbations of field [19] in the form $E = [E_a + e_1(t,z) + ie_2(t,z)]exp(-i\omega t + ikz)$, where $\omega$ is the frequency of a new branch, we linearize the Eq. (7), separate the real and imaginary parts, and obtain the equation set for $e_1$ and $e_2$

$$\frac{\partial^2 e_1}{\partial z^2} - 2k\frac{\partial e_2}{\partial z} - k^2 e_1 - \frac{\varepsilon_1'}{c^2}\left(\frac{\partial^2 e_1}{\partial t^2} + 2\omega\frac{\partial e_2}{\partial t} - \omega^2 e_1\right) + \frac{\varepsilon_1''}{c^2}\left(\frac{\partial^2 e_2}{\partial t^2} - 2\omega\frac{\partial e_1}{\partial t} - \omega^2 e_2\right)$$
$$- \frac{\chi' E_a^2}{c^2}\left(3\frac{\partial^2 e_1}{\partial t^2} + 2\omega\frac{\partial e_2}{\partial t} - 3\omega^2 e_1\right) + \frac{\chi'' E_a^2}{c^2}\left(\frac{\partial^2 e_2}{\partial t^2} - 6\omega\frac{\partial e_1}{\partial t} - \omega^2 e_2\right) = 0,$$
$$\frac{\partial^2 e_2}{\partial z^2} + 2k\frac{\partial e_1}{\partial z} - k^2 e_2 - \frac{\varepsilon_1'}{c^2}\left(\frac{\partial^2 e_2}{\partial t^2} - 2\omega\frac{\partial e_1}{\partial t} - \omega^2 e_2\right) - \frac{\varepsilon_1''}{c^2}\left(\frac{\partial^2 e_1}{\partial t^2} + 2\omega\frac{\partial e_2}{\partial t} - \omega^2 e_1\right)$$
$$- \frac{\chi' E_a^2}{c^2}\left(\frac{\partial^2 e_2}{\partial t^2} - 6\omega\frac{\partial e_1}{\partial t} - \omega^2 e_2\right) - \frac{\chi'' E_a^2}{c^2}\left(3\frac{\partial^2 e_1}{\partial t^2} + 2\omega\frac{\partial e_2}{\partial t} - 3\omega^2 e_1\right) = 0, \tag{8}$$

where $\chi = 4\pi\chi_{31}$. We represent the solutions of equation set (8) as $e_{1,2} = e_{01,02}exp(-i\Omega t + iKz)$ and obtain the dispersion relation for the phase additions $\Omega$ and $K$ from the determinant of set for $e_{01}$ and $e_{02}$,

$$K^4 - (4k^2 + a_1 + a_2)K^2 + 2k(b_1 + b_2)K + a_1 a_2 - b_1 b_2 = 0, \tag{9}$$

where

$a_1 = c^{-2}\left[(\omega^2 + \Omega^2)(\varepsilon_1' + 3\chi' E_a^2) + i2\omega\Omega(\varepsilon_1'' + 3\chi'' E_a^2)\right] - k^2$,

$a_2 = c^{-2}\left[(\omega^2 + \Omega^2)(\varepsilon_1' + \chi' E_a^2) + i2\omega\Omega(\varepsilon_1'' + \chi'' E_a^2)\right] - k^2$,

$b_1 = c^{-2}\left[2\omega\Omega(\varepsilon_1' + 3\chi' E_a^2) + i(\omega^2 + \Omega^2)(\varepsilon_1'' + 3\chi'' E_a^2)\right]$,

$b_2 = c^{-2}\left[2\omega\Omega(\varepsilon_1' + \chi' E_a^2) + i(\omega^2 + \Omega^2)(\varepsilon_1'' + \chi'' E_a^2)\right]$.

We can normalize the phase addition $\overline{K} = Kk^{-1}$ in the Eq. (9)

$$\overline{K}^4 - [4 + k^{-2}(a_1 + a_2)]\overline{K}^2 + 2k^{-2}(b_1 + b_2)\overline{K} + k^{-4}(a_1 a_2 - b_1 b_2) = 0. \tag{10}$$

We suppose the phase addition $\overline{K}$ is small enough $\overline{K}^4 \ll [4 + k^{-2}(a_1 + a_2)]\overline{K}^2$ and neglect the highest term. Then we obtain the quadratic equation from the Eq. (10)

$$\overline{K}^2 - 2a\overline{K} - b = 0, \tag{11}$$

where $a = (b_1 + b_2)(4k^2 + a_1 + a_2)^{-1}$, $b = k^{-2}(a_1 a_2 - b_1 b_2)(4k^2 + a_1 + a_2)^{-1}$. The instability of polariton wave with frequency $\omega$ and phase addition $\Omega$ takes place when the solutions of the Eq. (11)

$\overline{K}_\pm = a \pm (a^2 + b)^{1/2}$ have the complex values. The coefficients $a$ and $b$ depend on the field density $\sim E_a^2$ when the rest parameters are fixed. The dependence of real $Re\,\overline{K}_\pm$ and imaginary $Im\,\overline{K}_\pm$ parts of phase addition $\overline{K}_\pm$ on normalized field density $E_a^2$ for the polariton wave with normalized frequency $\overline{\omega} \cong 6.4$ is represented in the Fig. 2.

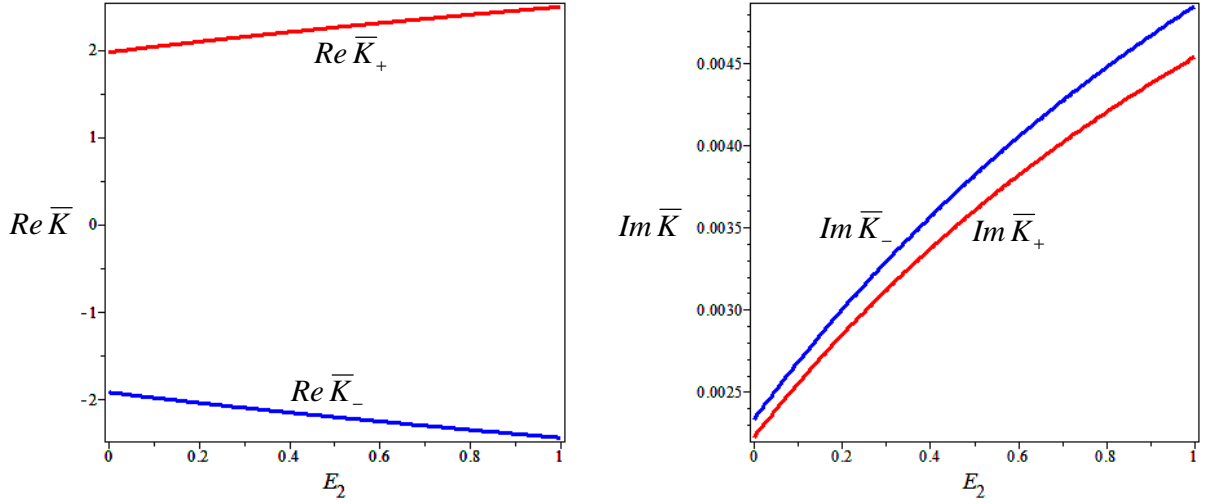

Fig. 2. The dependence of $Re\,\overline{K}_\pm$ and $Im\,\overline{K}_\pm$ on normalized field density $E_a^2$ for the polariton wave with normalized frequency $\overline{\omega} \cong 6.4$.

One can see in the Fig. 2 the both $Re\,\overline{K}_\pm$ and $Im\,\overline{K}_\pm$ are depended on the field density $E_a^2$. The imaginary parts $Im\,\overline{K}_\pm$ define the wave instability. If the imaginary parts of medium permittivity $\varepsilon_1''$ and susceptibility $\chi''$ are equal zero then $Im\,\overline{K}_\pm$ are equal zero too, and the wave instability is absent.

The perturbations $e_1$ and $e_2$ excite the modulation instability of nonlinear wave leading to the appearance of cnoidal waves and solitons [6, 7, 11, 12, 16 - 19]. We can represent the electric field of wave of a new spectrum branch in the form $E = e(t,z)exp(-i\omega t)$, then obtain from the Eq. (7) the equation for amplitude

$$\frac{\partial^2 e}{\partial z^2} - \frac{\varepsilon_1}{c^2}\left(\frac{\partial^2 e}{\partial t^2} - i2\omega\frac{\partial e}{\partial t} - \omega^2 e\right) - \\ \frac{4\pi\chi_{31}}{c^2}\left(e\frac{\partial^2|e|^2}{\partial t^2} + 2\frac{\partial|e|^2}{\partial t}\frac{\partial e}{\partial t} + |e|^2\frac{\partial^2 e}{\partial t^2} - i2\omega e\frac{\partial|e|^2}{\partial t} - i2\omega|e|^2\frac{\partial e}{\partial t} - \omega^2|e|^2 e\right) = 0. \quad (12)$$

In the case of stationary process of the wave propagation we can suppose that $e(t,z)$ is the slowly varying at time amplitude and reduce the Eq. (12)

$$\frac{\partial^2 e}{\partial z^2} + \frac{\omega^2}{c^2}\left(\varepsilon_1 + 4\pi\chi_{31}|e|^2\right)e = 0. \tag{13}$$

The first integral of the Eq. (13) looks like $(de/dz)^2 = -\bar{\varepsilon}e^2 - \bar{\chi}e^4/2 + C$, where $C = (de/dz)_0^2 + \bar{\varepsilon}e_0^2 + \bar{\chi}e_0^4/2$ is an integration constant, $\bar{\varepsilon} = c^{-2}\omega^2\varepsilon_1$, $\bar{\chi} = 4\pi c^{-2}\omega^2\chi_{31}$. For wave of new spectrum branch the boundary condition is $e_0 = 0$ and $C = (de/dz)_0^2$. Thus, the second integral of Eq. (13) looks like $\sqrt{\bar{\chi}/2}\,z = \int de\left(C' - \alpha e^2 - e^4\right)^{-1/2}$, where $C' = 2C/\bar{\chi}$, $\alpha = 2\bar{\varepsilon}/\bar{\chi}$. After integration of the second integral we obtain the solution in the form of elliptic cosine

$$e = B\,cn\left\{K(\tilde{k}) - \left[\bar{\chi}\left(\alpha^2/4 + C'\right)^{1/2}\right]^{1/2} z, \tilde{k}\right\}, \tag{14}$$

where $K(\tilde{k})$ is the complete elliptic integral, $\tilde{k} = B/(A^2 + B^2)^{1/2}$ is the modulus of elliptic integral, $A^2 = \alpha/2 + (\alpha^2/4 + C')^{1/2}$, $B^2 = -\alpha/2 + (\alpha^2/4 + C')^{1/2}$. In the case when the modulus tend to unit $\tilde{k} \to 1$ (if $\alpha < 0$ and $C' \to 0$), the Eq. (14) transforms to hyperbolic secant describing a spatial soliton

$$e = |\sqrt{\alpha}|\,sch(\sqrt{\bar{\varepsilon}}\,z). \tag{15}$$

## 6. CONCLUSION

We have shown theoretically that the dispersion dependence of the third order susceptibility leads to appearance of the new branches in polariton spectrum at nonlinear dielectric medium. Particularly, the new branches appear in the polariton spectrum gap, when the electromagnetic field density increases. The high frequency polaritons decay to three low frequency polaritons under the influence of the third order susceptibility of dielectric medium. The polariton waves of new spectrum branches damp more rapidly, because they depend of nonlinear medium properties and the wave intensity. The instability of polariton wave depends on the field density. The perturbations excite the nonlinear wave modulation instability leading to appearance of cnoidal waves and solitons.

These properties of the polariton spectrum in nonlinear dielectric medium may be used for designing new optical devices such as controllable optical filters, delay lines, all-optical logic gates, etc.

**ACKNOWLEDGMENT**

The authors are grateful to Anton S. Desyatnikov for fruitful discussions about the paper.**REFERENCES**

1. Tolpygo K. B., "Physical properties of the salt lattice constructed from deforming ions", *JETP*, No. 6, **20**, 497-509 (1950) (in Russian).

2. Huang K., "On the interaction between the radiation field and ionic crystals", *Proc Roy. Soc. A*, 352-365 (1951).

3. Agranovich V. M. and Ginzburg V. L., *Crystal optics with spatial dispersion and excitons* (Springer, 1984).

4. Albuquerque E. L., Cottam M. G., *Polaritons in periodic and quasiperiodic structures* (Elsevier, Amsterdam, Netherlands, 2004).

5. Klyshko D. N., *Quantum and nonlinear optics* (Nauka, 1980) (in Russian).

6. Shen Y. R., *The principles of nonlinear optics* (John Wiley & Sons, 1984).

7. Dzedolik I. V., *Polaritons in optical fibers and dielectric resonators* (DIP, 2007) (in Russian).

8. Baher S., Cottam M. G., "Theory of Nonlinear s-Polarized Phonon-Polaritons in Multilayered Structures", Journal of Sciences, Islamic Republic of Iran, **15(2)**, 171-177 (2004).

9. Inoue H., Katayma K., Shen Q., Toyoda T., Nelson K., "Terahertz reflection response measurement using a phonon polariton wave", *J. Appl. Phys.*, **105,** 054902 (2009).

10. Qi Z., Shen Z-Q., Huang C-P., Zhu S-N., Zhu Y-Y., "Phonon polaritons in a nonaxial aligned piezoelectric superlattice", *J. Appl. Phys.*, **105,** 074102 (2009).

11. Dzedolik I. V., "Period variation of polariton waves in optical fiber", *J. Opt. A: Pure Appl. Opt.*, **11,** 094012 (2009).

12. Dzedolik I. V., Lapayeva S. N., "Mass of polaritons in different dielectric media", *J. Opt.*, **13,** 015204 (2011).

13. Dzedolik I. V., Karakchieva O. S., "Polaritons in nonlinear medium: generation, propagation and interaction", *Proc. NLP*2011 IEEE*, Catalog Number: CFP1112P-CDR, ISBN: 978-1-4577-0479-6, 3 p. (2011).

14. Campbell G., Hosseini M., Sparkes B. M., Lam P. K., Buchler B. C., "Time- and frequency-domain polariton interference", *New J. Phys.*, **14,** 033022 (2012).

15. Kudryashov N. A., Ryabov P. N., Sinelshchikov D. I., "Nonlinear waves in media with fifth order dispersion", *Opt. Lett. A*, **375**, Is. 20, P. 2051–2055 (2011).

16. Gaizauskas E., Savickas A., Staliunas K., "Radiation from band-gap solitons", *Opt. Com.*, **285**, Is. 8, P. 2166–2170 (2012).
10